\documentclass[10pt,superscriptaddress,twocolumn,showpacs,
amssymb,amsmath,nobibnotes,aps,prd,
nofootinbib,floatfix]{revtex4-1}
\pdfoutput=1

\usepackage{graphicx,bm,color,psfrag}
\usepackage[caption=false]{subfig}
\usepackage{amsfonts}
\usepackage{lipsum}
\usepackage{mathtools}
\usepackage{verbatim}
\usepackage[normalem]{ulem}
\usepackage[dvipsnames]{xcolor}
\usepackage{dcolumn}
\usepackage{latexsym}
\usepackage{mathrsfs}
\usepackage{booktabs}

\usepackage{hyperref} % SEMPRE por último
\hypersetup{colorlinks,linkcolor={blue},citecolor={red},urlcolor={greenW}} 
\definecolor{greenW}{rgb}{0.0, 0.55, 0.1}

\newcommand{\be}{\begin{equation}}
\newcommand{\ee}{\end{equation}}
\newcommand{\bea}{\begin{eqnarray}}
\newcommand{\eea}{\end{eqnarray}}

\begin{document}

\title{Tracing the Evolution of \texorpdfstring{$\Omega_m(z)$}{Omegam(z)} over the Last 10 Billion Years with Non-parametric Methods}

\author{R. F. L. Holanda}
\email{holandarfl@gmail.com}
\affiliation{Universidade Federal do Rio Grande do Norte, Departamento de F\'isica Te\'orica e Experimental, 59300-000, Natal - RN, Brazil}

\author{J. F. Jesus}
\email{jf.jesus@unesp.br}
\affiliation{Universidade Estadual Paulista (UNESP), Instituto de Ci\^encias e Engenharia, Itapeva, SP, Brazil}
\affiliation{Universidade Estadual Paulista (UNESP), Instituto de Geoci\^encias e Ci\^encias Exatas, Rio Claro, SP, Brazil}
\affiliation{Universidade Estadual Paulista (UNESP), Faculdade de Engenharia e Ci\^encias, Guaratinguet\'a, SP, Brazil}

\author{Z. C. Santana}
\email{zilmarcsjunior@gmail.com}
\affiliation{Universidade Federal do Rio Grande do Norte, Departamento de F\'isica Te\'orica e Experimental, 59300-000, Natal - RN, Brazil}

\author{R. C. Nunes}
\email{rafadcnunes@gmail.com}
\affiliation{Instituto de F\'isica, Universidade Federal do Rio Grande do Sul, 91501-970 Porto Alegre, RS, Brazil}
\affiliation{Divis\~ao de Astrof\'isica, Instituto Nacional de Pesquisas Espaciais, S\~ao Jos\'e dos Campos, SP, Brazil}

\begin{abstract}
We investigate the redshift evolution of the matter density parameter, $\Omega_m(z)$, using galaxy cluster gas mass fraction measurements combined with cosmic chronometer $H(z)$ data and type Ia supernova luminosity distances. This provides a non-parametric probe of the normalization of the cosmological matter sector, which plays a central role in current tensions involving weak lensing measurements, cluster abundance analyses, and the $S_8$ parameter. Using Gaussian Process regression, we reconstruct $\Omega_m(z)$ without assuming a parametric form for its evolution. The reconstructed evolution is consistent with the standard $\rho_m \propto (1+z)^3$ scaling predicted by the $\Lambda$CDM model. We obtain $\Omega_{m0}=0.296 \pm 0.044$ from the 44-cluster sample, and $\Omega_{m0}=0.271 \pm 0.016$, $0.253 \pm 0.017$, and $0.210 \pm 0.013$ for the 103-cluster compilation, depending on the adopted mass calibration. While the reconstructed $\Omega_m(z)$ evolution remains consistent with the expected $\Lambda$CDM behaviour, the inferred normalization of $\Omega_{m0}$ depends strongly on the adopted cluster mass calibration. Consequently, cluster mass calibration systematics constitute the dominant source of uncertainty in the inferred normalization of $\Omega_{m0}$, exceeding the statistical uncertainties of the non-parametric reconstruction.
\end{abstract}

\maketitle

\section{Introduction}

Galaxy clusters are the most massive gravitationally bound structures in the Universe, typically residing at the nodes of the cosmic web \cite{Miyatake:2025iqa}. With virial masses in the range $10^{14}$--$10^{15}\,M_{\odot}$, they are composed predominantly of dark matter ($\sim 80$--$85\%$), while the remaining mass is distributed between the hot intracluster medium (ICM) and the stellar component associated with cluster galaxies. Formed from the collapse of rare high-density peaks in the primordial matter distribution, clusters grow hierarchically through accretion and mergers, thereby tracing the evolution of large-scale structure. Owing to their large masses and relatively well-understood gravitational physics, galaxy clusters have long been recognized as powerful laboratories for both astrophysics and cosmology (see, e.g., \cite{allen2011cosmological,kravtsov2012formation,pratt2019}).

The cosmological relevance of galaxy clusters has been established for decades, and in the era of precision cosmology they have become key probes of structure growth. Measurements of cluster abundance and the mass function have been widely used to constrain the matter density parameter $\Omega_{m0}$ and the amplitude of matter fluctuations $\sigma_8$ \cite{Allen2011,Kravtsov2012,Planck2016Clusters,DES2020Clusters,SPT2022Clusters,2025APh...16503052S,2022EPJC...82...17B,2021EPJC...81..596B,2019JCAP...11..032H}. Large samples identified through X-ray, optical, and Sunyaev--Zel'dovich surveys---including those from the South Pole Telescope, the Atacama Cosmology Telescope, the Dark Energy Survey, and more recently the eROSITA mission---have significantly improved the statistical power of cluster-based cosmology \cite{Zhong:2026pwh,Allen2011,Kravtsov2012,DES2020Clusters,SPT2022Clusters,ACT2023Clusters,eROSITA2017,Desjacques2018LSS,Lau:2022dub}. Interestingly, several analyses report values of the fluctuation amplitude that are slightly lower than those inferred from Cosmic Microwave Background observations, contributing to the ongoing discussion of the $S_8$ tension (see \cite{CosmoVerseNetwork:2025alb,Pantos:2026koc} for a review). These results highlight the importance of independent probes capable of testing the robustness of the standard $\Lambda$CDM model \cite{CosmoVerseNetwork:2025alb}. In this context, determining the normalization of the matter sector has become increasingly relevant, particularly due to current tensions involving weak lensing, cluster abundance measurements, and the $S_8$ parameter. While measurements of $H(z)$ probe the background expansion, cluster baryon observables such as the gas mass fraction provide complementary information on the relation between baryonic and total matter components, allowing consistency tests of the inferred matter density normalization.

Despite their strong potential, the use of galaxy clusters as precision cosmological probes is limited by the difficulty of accurately determining their total masses. Weak gravitational lensing provides the most robust method for mass calibration, but such measurements remain observationally demanding and are available only for relatively limited samples \cite{pratt2019,umetsu2020,wu2022,euclid2025}. Alternative approaches based on galaxy dynamics or virial analyses require extensive spectroscopic data and are subject to systematic uncertainties related to projection effects and the dynamical state of clusters. In addition, X-ray mass estimates derived under the assumption of hydrostatic equilibrium may be biased due to non-thermal pressure support, leading to the so-called hydrostatic mass bias \cite{Old2020,Saro2022,Angelinelli2020,Ettori2022,Braspenning2024}.

Given these challenges, cosmological tests based on internal cluster properties provide a valuable complementary approach. One well-established observable is the cluster gas mass fraction, $f_{\rm gas}$, defined as the ratio between the baryonic gas mass and the total mass of the cluster. Under the assumption that massive, relaxed clusters provide a fair sample of the cosmic baryon fraction, measurements of $f_{\rm gas}$ have been widely used to constrain cosmological parameters and to test possible deviations from the standard cosmological model (e.g., \cite{Allen2008,Allen2011,Mantz2014,Mantz2015,10.1093/mnras/stab3390,Holanda2010,Holanda2011,Holanda2012,Holanda2016,Holanda2017,Gonzalez2024}). In addition to their use as cosmological probes, measurements of the cluster gas mass fraction provide a useful framework for testing scenarios in which the matter component may deviate from the standard evolution law expected in the $\Lambda$CDM model. Such tests can be formulated in a largely model-independent way and are often described through phenomenological parametrizations of the matter density, typically written as $\rho_m=\rho_{m,0}(1+z)^{3+\epsilon}$, where $\epsilon$ quantifies departures from the standard $\rho_m\propto(1+z)^3$ scaling \cite{Wang:2004cp, Bora:2021iww, Holanda:2019sod, santana2024interaction}.  Deviations from this standard behaviour are commonly motivated by extensions of the $\Lambda$CDM framework, such as models involving non-gravitational interactions in the dark sector, in which dark matter and dark energy coevolve. These scenarios have been explored as possible resolutions to fundamental problems in cosmology, including the coincidence problem and the cosmological constant problem, and are also connected to approaches that consider a dynamical vacuum energy within the context of quantum field theory~\cite{Linde:1974at, Nelson:1982kt, polyakov1982phase, Weinberg:1988cp, Elizalde:1993ew, Bytsenko:1994at, Elizalde:1995at,  Cohen:1998zx,   Shapiro:1999zt, shapiro2002scaling, Shapiro:2003ui, Sola:2011qr}.

In this work we adopt a { non-parametric} approach to investigate the redshift evolution of the matter density parameter $\Omega_m(z)$. Our methodology does not rely on the assumption of a specific cosmological model nor on a parametric description of the matter evolution, allowing a direct reconstruction of cosmological quantities from observational data. To this end, we combine two galaxy cluster gas mass fraction samples with cosmic chronometer measurements of $H(z)$ and type Ia supernova luminosity distances, enabling us to estimate both $\Omega_m(z)$ and its present-day value $\Omega_{m0}$. The reconstruction is performed using Gaussian Process Regression (GPR), a non-parametric technique that has been widely applied in cosmology, including studies of dark sector dynamics \cite{holsclaw2010nonparametric,seikel2012reconstruction,yahya2014null,santana2024interaction,santana2025non,yang2025gaussian}, the $H_0$ tension \cite{busti2014evidence,gomez2018h,liao2019model,colacco2025joint,colacco2025hubble}, tests of fundamental cosmological relations \cite{santos2015two,ruan2018model,gong2024multiple,luo2025testing}, calibration of astrophysical probes \cite{holanda2017cosmological,zhang2024constraints}, and statistical inference frameworks \cite{johnson2025kernel,mukherjee2025revisiting}.  In Section~\ref{sec:method}, we present the methodology. The data analysis and results are discussed in Section~\ref{sec:analysis}. Finally, we summarize our conclusions in Section~\ref{sec:conclusion}.

\section{\label{sec:method}Methodology}

In this section, we present the theoretical framework and practical implementation adopted to infer the redshift-dependent matter density parameter $\Omega_m(z)$ by combining galaxy cluster gas mass fraction measurements with $H(z)$ data and type Ia supernova (SNe Ia) observations.

We follow the modeling of the cluster gas mass fraction, $f_{\rm gas}$, as described in \cite{SPT:2021vsu}, where $f_{\rm gas}$ is allowed to depend on cluster mass through a power-law relation with slope $\alpha$. In this framework, the observed gas mass fraction, accounting for systematic effects and the use of a reference cosmology, can be written as
\begin{equation}
    f_{\rm gas}(z) = \Upsilon(z)\,K(z)\frac{\Omega_b(z)}{\Omega_m(z)}
    \left[\frac{d^\star_L(z)}{d_L(z)}\right]^{3/2},
    \label{eq:f_def}
\end{equation}
{ where $\Upsilon(z)$ denotes the depletion factor. Depending on the observational definition adopted for the cluster baryon fraction, this quantity may describe the depletion of the gas component alone or of the total baryonic component. The specific definition depends on the observational methodology adopted for each cluster sample and is discussed in the corresponding subsections below.} $K(z)$ accounts for possible redshift-dependent biases arising from systematic uncertainties in total mass estimates. The quantities $\Omega_b(z)$ and $\Omega_m(z)$ are the baryon and total matter density parameters, respectively. The term $d^\star_L(z)$ corresponds to the luminosity distance in a fiducial cosmology\footnote{We adopt a spatially flat $\Lambda$CDM fiducial model with $H_0=70~{\rm km\,s^{-1}Mpc^{-1}}$ and $\Omega_{m0}=0.3$.}, while $d_L(z)$ is the luminosity distance inferred from SNe Ia observations. { The dependence on the fiducial cosmology adopted in the original X-ray data reduction is explicitly accounted for by the factor $(d_L^*/d_L)^{3/2}$, which replaces the fiducial luminosity distance with the reconstructed one. This treatment, widely adopted in cosmological analyses based on the cluster gas mass fraction (e.g., \cite{Allen2008,Allen2011,Mantz2014,Holanda2017}), ensures that the cosmological inference is not tied to the specific fiducial cosmology assumed in the original data reduction.}

The critical density at redshift $z$ is defined as
\begin{equation}
    \rho_c(z) = \frac{3H^2(z)}{8\pi G},
\end{equation}
and the baryon density evolves as
\begin{equation}
    \rho_b(z) = \rho_{b0}(1+z)^3,
\end{equation}
assuming standard conservation of baryon number.

It is therefore convenient to express the baryon density parameter as
\begin{equation}
    \Omega_b(z)
    = \frac{\rho_b(z)}{\rho_c(z)}
    = \frac{\Omega_{b0}(1+z)^3 H_0^2}{H^2(z)},
    \label{eq_Omega_bz}
\end{equation}
where $\Omega_{b0} \equiv \rho_{b0}/\rho_{c0}$ and $\rho_{c0}=3H_0^2/(8\pi G)$.

By combining Eq.~\eqref{eq:f_def} with Eq.~\eqref{eq_Omega_bz}, we obtain an expression for the redshift-dependent matter density parameter:
\begin{equation}
    \Omega_m(z)=
    \frac{\Omega_{b0}(1+z)^3 H_0^2}{H^2(z)}
    \frac{\Upsilon(z)K(z)}{f_{\rm gas}(z)}
    \left(\frac{d^\star_L}{d_L}\right)^{3/2}.
    \label{eq:Omega_mz}
\end{equation}

This relation provides a direct way to reconstruct $\Omega_m(z)$ from observational quantities. In particular, $H(z)$ measurements from cosmic chronometers enter through the critical density, while SNe Ia data provide the luminosity distance $d_L(z)$. The gas mass fraction measurements act as the primary tracer linking baryonic and total matter components.

{ To reconstruct $\Omega_m(z)$ in a non-parametric manner, we employ Gaussian Process Regression (GPR) \cite{williams2006gaussian,seikel2012reconstruction}. This approach allows us to infer smooth functions from discrete data sets without assuming a specific functional form for their redshift dependence. In practice, GPR is applied to reconstruct $H(z)$ and $d_L(z)$ from observational data, which are then propagated through Eq.~\eqref{eq:Omega_mz} to obtain $\Omega_m(z)$.}

Gaussian Processes (GPs) provide a non-parametric method for reconstructing 
a function $f(x)$ from a set of measurements $f(x_i)\pm\sigma_i$. A GP 
assumes that the function values evaluated at any finite set of points follow 
a multivariate Gaussian distribution. It is completely specified by its mean 
function and covariance function (or kernel),
\begin{align}
\mu(x)&=\langle f(x)\rangle,\\
k(x,x')&=\left\langle
[f(x)-\mu(x)][f(x')-\mu(x')]
\right\rangle,\\
\mathrm{Var}[f(x)]&=k(x,x).
\end{align}
Therefore, a GP is denoted by
\be
f(x)\sim\mathcal{GP}\bigl(\mu(x),k(x,x')\bigr).
\ee

Unlike parametric reconstructions, the GP method does not assume a specific 
functional form for $f(x)$. Nevertheless, the reconstruction depends on the 
choice of kernel, which determines how the function values at different input 
points are correlated. In general, this correlation decreases as the distance 
$|x-x'|$ increases. We adopt the commonly used Squared Exponential kernel,
\be
k(x,x')=\sigma_f^2
\exp\left[-\frac{(x-x')^2}{2l^2}\right],
\ee
{where $\sigma_f$ controls the typical amplitude of the function variations, 
whereas $l$ determines their characteristic correlation length. Although 
other kernels, such as the Mat\'ern family, are also frequently used in 
cosmology, previous analyses \cite{Yu:2017iju,Jesus:2019nnk} have shown that the resulting reconstructions 
are often only weakly dependent on this choice. On the other hand, since the uncertainties in our reconstructions are typically large, we find that the results obtained with different kernel choices are mutually consistent. We therefore restrict our analysis to the Squared Exponential kernel.}

{The hyperparameters $\sigma_f$ and $l$ are determined by maximizing the log 
marginal likelihood \cite{seikel2012reconstruction},}
\begin{align}
\ln\mathcal{L}
&=\ln p(\mathbf{y}\mid\mathbf{X},\sigma_f,l)\nonumber\\
&=-\frac{1}{2}(\mathbf{y}-\boldsymbol{\mu})^T
[K(\mathbf{X},\mathbf{X})+C]^{-1}
(\mathbf{y}-\boldsymbol{\mu})\nonumber\\
&\quad-\frac{1}{2}
\ln\left|K(\mathbf{X},\mathbf{X})+C\right|
-\frac{n}{2}\ln(2\pi),
\end{align}
{where $\mathbf{X}=\{x_i\}$ is the set of input points, $\mathbf{y}$ is the 
data vector, $\boldsymbol{\mu}$ is the mean vector, $C$ is the data covariance 
matrix, and $n$ is the number of measurements. The components of the kernel 
matrix are}
\be
K_{ij}(\mathbf{X},\mathbf{X})=k(x_i,x_j).
\ee

{Once the hyperparameters have been determined, the function can be 
reconstructed at a set of points $\mathbf{X}^*$ according to}
\be
\left\langle\mathbf{f}^*\right\rangle
=\boldsymbol{\mu}^*
+K(\mathbf{X}^*,\mathbf{X})
[K(\mathbf{X},\mathbf{X})+C]^{-1}
(\mathbf{y}-\boldsymbol{\mu}),
\label{frec}
\ee
{with covariance}
\begin{align}
\operatorname{cov}(\mathbf{f}^*)
={}&K(\mathbf{X}^*,\mathbf{X}^*)\nonumber\\
&-K(\mathbf{X}^*,\mathbf{X})
[K(\mathbf{X},\mathbf{X})+C]^{-1}
K(\mathbf{X},\mathbf{X}^*).
\label{covfrec}
\end{align}

{ Although the GP reconstruction is non-parametric with respect to the redshift dependence of the reconstructed functions, the analysis is not strictly model independent. It relies on assumptions such as standard baryon conservation, a prior on $\Omega_{b0}h^2$, a fiducial cosmology used in the original cluster data reduction, and simulation- or observation-based prescriptions for the depletion and mass-calibration factors. This strategy minimizes the dependence on a particular cosmological model and enables a direct comparison between the reconstructed $\Omega_m(z)$ and the evolution implied by the standard matter-density scaling $\rho_m(z)=\rho_{m0}(1+z)^3$. Since $\Omega_m(z)=\rho_m(z)/\rho_c(z)$, the $\Lambda$CDM expectation is not a constant function, but rather
\begin{equation}
\Omega_m(z)=\frac{\Omega_{m0}(1+z)^3}{E^2(z)},
\end{equation}
where $E(z)=H(z)/H_0$. Accordingly, deviations from the standard matter-density scaling would appear as departures of the reconstructed $\Omega_m(z)$ from the redshift evolution predicted by the $\Lambda$CDM framework.}

{ Our analysis relies on several key assumptions. First, the baryon density follows the standard evolution $\rho_b \propto (1+z)^3$. Second, the depletion factor $\Upsilon(z)$ and calibration term $K(z)$ are assumed to evolve only slowly with redshift, as suggested by hydrodynamical simulations and previous observational analyses \cite{planelles2013cluster,planelles2013baryon,allen2011cosmological,Mantz2014,Mantz2015,Mantz2022}. The specific implementation adopted for these quantities is described in the following section. Residual uncertainties associated with them are treated as systematic errors and propagated through the reconstruction.}
As discussed above, the use of a fiducial cosmology in $d_L^\star(z)$ does not introduce significant bias, as its effect is explicitly corrected by the ratio $(d^\star_L/d_L)^{3/2}$. Equation~\eqref{eq:Omega_mz} is the basic working formula used to derive $\Omega_m(z)$ from the cluster observables. The parametrizations for the nuisance functions $\Upsilon(z)$ and $K(z)$ will be specified according to each cluster sample, as explained below.

\begin{figure*}[t]
    \centering
    \includegraphics[width=0.49\textwidth]{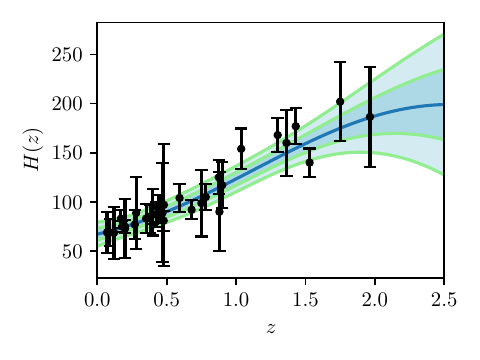}
    \includegraphics[width=0.49\textwidth]{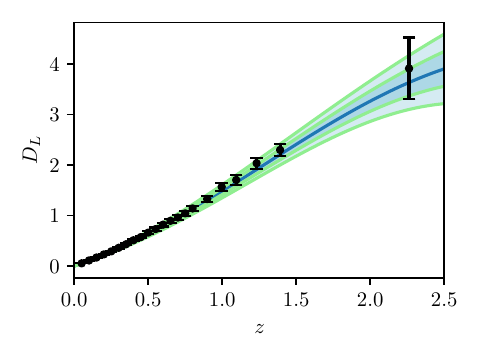}
\caption{Gaussian Process reconstructions of cosmological observables. Left: $H(z)$ reconstructed from 32 cosmic chronometer measurements \cite{Moresco:2022phi}. Right: luminosity distance $D_L(z)$ reconstructed from the binned Union3 dataset \cite{Rubin:2023jdq}. Shaded regions represent the $1\sigma$ and $2\sigma$ confidence intervals.}
    \label{fig:HzSNeData}
\end{figure*}

\section{Data sets and Results}
\label{sec:analysis}

To reconstruct the continuous evolution of $\Omega_m(z)$ using GPR, we first compute its discrete estimates by evaluating all terms on the right-hand side of Eq.~\eqref{eq:Omega_mz}. This requires independent reconstructions of the kinematic quantities $H(z)$ and $d_L(z)$.

{For the Hubble parameter, we use direct measurements from cosmic chronometers (CC), compiled in Ref.~\cite{Moresco:2022phi} (see Table 1 therein). The CC compilation consists of 32 measurements, including both statistical and systematic uncertainties. It is important to note that CC measurements may be affected by systematic uncertainties associated with the determination of the age of the Universe, the stellar population synthesis models, and the assumptions involved in estimating galaxy ages. The resulting reconstruction is shown in the left panel of Fig.~\ref{fig:HzSNeData}.}

%The Hubble parameter $H(z)$ is obtained through a Gaussian Process reconstruction of cosmic chronometer (CC) data \cite{williams2006gaussian,seikel2012reconstruction,Moresco:2022phi}. The CC compilation consists of 32 measurements of $H(z)$, including both statistical and systematic uncertainties. The resulting reconstruction is shown in the left panel of Fig.~\ref{fig:HzSNeData}.

For the luminosity distance, we use the Union3 compilation \cite{Rubin:2023jdq}, which provides 22 binned apparent magnitude measurements derived from 2087 type Ia supernovae (SNe Ia). However, as discussed in \cite{Jesus:2019nnk}, the apparent magnitude $m(z)$ exhibits a highly nonlinear behavior at low redshift ($m \rightarrow -\infty$ as $z \rightarrow 0$), making it unsuitable for direct GP reconstruction. Following \cite{Jesus:2019nnk}, we first propagate the uncertainties to obtain 22 measurements of the dimensionless luminosity distance, defined as $D_L \equiv H_0 d_L / c$. We then reconstruct $D_L(z)$ using GPR. {The dimensionless $D_L(z)$ was obtained by using the fact that the GPR reconstructs not only $d_L(z)$, but also its derivative, $d_L'(z)$. Recalling that all cosmological distances scales as $d_L(z)\simeq\frac{cz}{H_0}$ at low $z$, it means that $d_L'(z)\simeq\frac{c}{H_0}$ at low $z$. So we have used the $d_L'(z)$ reconstruction to write the dimensionless luminosity distance as $D_L(z)=\frac{d_L(z)}{d_L'(0)}$.} The resulting reconstruction is shown in the right panel of Fig.~\ref{fig:HzSNeData}.

We note that the ratio entering Eq.~\eqref{eq:Omega_mz} satisfies
\begin{equation}
    \frac{d_L^\star}{d_L} = \frac{D_L^\star}{D_L},
\end{equation}
which ensures that the analysis is independent of the absolute calibration of $H_0$.

%\textcolor{blue!70!black}{Referee (add new text here): The construction of the dimensionless luminosity distance, $D_L = H_0 d_L/c$, needs clarification. The authors should specify the value of $H_0$ used in this conversion. If the supernova binned data or the adopted conversion assumes $H_0 = 70\,{\rm km\,s^{-1}\,Mpc^{-1}}$, this should be stated explicitly.}
%\\

%\textcolor{blue!70!black}{Referee (add text here): A short discussion of these systematics (CC, $f_gas$ and SNIa), and how they may
%affect the final $\Omega_{m}(z)$ reconstruction would improve the clarity of the analysis.}
%\\

To evaluate the present-day baryon density parameter $\Omega_{b0}$, which acts as a normalization factor in Eq.~\eqref{eq:Omega_mz}, we adopt the prior constraint \cite{hsyu2020phlek}
\begin{equation}
    \Omega_{b0}h^2 = 0.0215 \pm 0.0005.
\end{equation}
This value is based on recent determinations of the primordial helium abundance from observations of metal-poor star-forming galaxies. In particular, near-infrared spectroscopy of the He\,I $\lambda10830$~\AA\ emission line, combined with optical data, yields a primordial helium mass fraction $Y_P = 0.2436^{+0.0040}_{-0.0039}$. When combined with the latest primordial deuterium measurements, this leads to the above constraint on $\Omega_{b0} h^2$, which is consistent within $1.3\sigma$ with the value inferred from Planck observations of the Cosmic Microwave Background. {We verified that adopting different reasonable priors on $\Omega_{b0}h^2$ does not affect our main results or their interpretation.}

\begin{figure*}[t]
    \centering
    \includegraphics[width=0.49\textwidth]{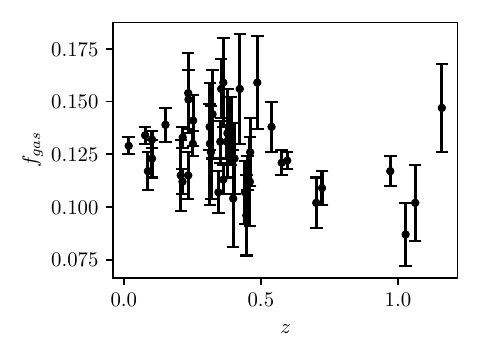}
    \includegraphics[width=0.49\textwidth]{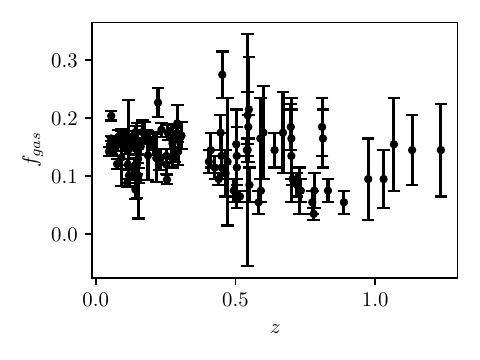}
    \caption{Gas mass fraction ($f_{\rm gas}$) measurements for the cluster samples. The left panel corresponds to the 44-cluster sample, while the right panel shows the 103-cluster sample.}
    \label{fig:44cldata}
\end{figure*}

{In short, the methodology adopted here relies on a set of well-defined assumptions and fiducial inputs. We adopt a flat $\Lambda$CDM cosmology as reference, with $H_0 = 70\text{ km  s}^{-1}\text{Mpc}^{-1}$ and $\Omega_{m0} = 0.3$, which enters through the fiducial luminosity distance $d_L^\star(z)$. Any dependence on this choice is explicitly corrected by the factor $(d_L^\star/d_L)^{3/2}$. We assume standard baryon conservation, such that $\rho_b(z) \propto (1+z)^3$, implying a specific redshift evolution for $\Omega_b(z)$ given $\Omega_{b0}$ and $H_0$.} The reconstruction is based on observational inputs from galaxy cluster gas mass fraction measurements, $H(z)$ data from cosmic chronometers, and luminosity distances from type Ia supernovae. Astrophysical systematics are encoded in the depletion factor $\Upsilon(z)$ and calibration term $K(z)$, which are assumed to be constant or slowly varying with redshift, with associated uncertainties treated as systematics. Finally, GPR is employed to reconstruct $H(z)$ and $d_L(z)$ in a non-parametric way, minimizing model dependence while allowing a direct test of the standard scaling $\rho_m \propto (1+z)^3$. 

{It is important to mention that although the GP methodology circumvents the need to impose a specific parametric model for $\Omega_m(z)$, it does not render the reconstruction ``model-independent" in an absolute sense: each of the three probes carries its own chain of astrophysical (cluster physics, stellar population modeling, SNIa standardization) and cosmological (distance duality, fiducial cosmology) assumptions. The reconstruction should therefore be interpreted as a joint consistency test of these datasets and their underlying systematics, rather than as a probe-independent measurement of $\Omega_m(z)$. An unidentified systematic bias in any single dataset would propagate through the GP and remain observationally indistinguishable from genuine cosmological evolution.}

\subsection{Sample of 44 clusters}

The first dataset, compiled by Mantz et al.~(2022) \cite{10.1093/mnras/stab3390}, consists of 44 gas mass fraction measurements spanning the redshift range $0.078 < z < 1.063$. These measurements are extracted from spherical shells within the radial interval $0.8$--$1.2\,r_{2500c}$, where $r_{2500c}$ denotes the radius enclosing a mean density equal to 2500 times the critical density of the Universe. 

To ensure robust cosmological constraints, the sample is restricted to the most massive and dynamically relaxed clusters observed with the \textit{Chandra} X-ray Observatory. Measuring $f_{\rm gas}$ in this intermediate radial region avoids the cluster core, thereby reducing systematic uncertainties associated with baryonic processes such as radiative cooling and feedback from active galactic nuclei. The dataset used in this analysis is shown in the left panel of Fig.~\ref{fig:44cldata}.

Here $M_{2500}$ denotes the total mass enclosed within $r_{2500c}$, and the exponent $\alpha = 0.025 \pm 0.033$ characterizes the mass dependence of $f_{\rm gas}$. The parameters $\Upsilon_0$ and $\Upsilon_1$ correspond to the normalization of the gas depletion factor and its possible redshift evolution, respectively. Since the gas mass fraction in this sample is measured within a spherical shell around $r_{2500}$, the depletion factor refers exclusively to the hot intracluster gas. {We adopt the Gaussian priors $\Upsilon_0 = 0.79 \pm 0.14$ and $\Upsilon_1 = 0.07 \pm 0.12$}, motivated by hydrodynamical simulations that include radiative cooling, star formation, and feedback from both active galactic nuclei and supernovae \cite{planelles2013baryon,planelles2013cluster}.

We also assume a redshift-dependent parametrization for the calibration factor, $K(z)=K_{0}(1+K_{1}z)$, where $K_0$ represents the overall normalization and $K_1$ allows for a possible redshift evolution of the calibration bias. {The normalization is constrained by the Gaussian prior $K_{0}=0.93 \pm 0.11$, derived from a subsample of clusters with independent weak-lensing mass measurements. The evolution parameter is restricted to the interval $-0.05 \leq K_{1} \leq 0.05$ \cite{planelles2013cluster,planelles2013baryon,10.1093/mnras/stab3390}.}

To derive $\Omega_m(z)$ from the $f_{\rm gas}(z)$ measurements, we use Eq.~\eqref{eq:Omega_mz}. The functions $\Upsilon(z)$ and $K(z)$ are computed using the parametrizations above, propagating the uncertainties in the parameters $\Upsilon_0$, $\Upsilon_1$, $\alpha$, {and $K_0$. Since $K_1$ is assumed to follow a uniform distribution in the interval $[-0.05,\,0.05]$, we approximate its uncertainty by an equivalent Gaussian standard deviation, $\sigma_{K_1}=0.05/\sqrt{3}\approx 0.029$.} This approximation enables a consistent treatment of uncertainties within the standard error propagation framework.
\begin{figure}[t]
    \centering
    \includegraphics[width=.49\textwidth]{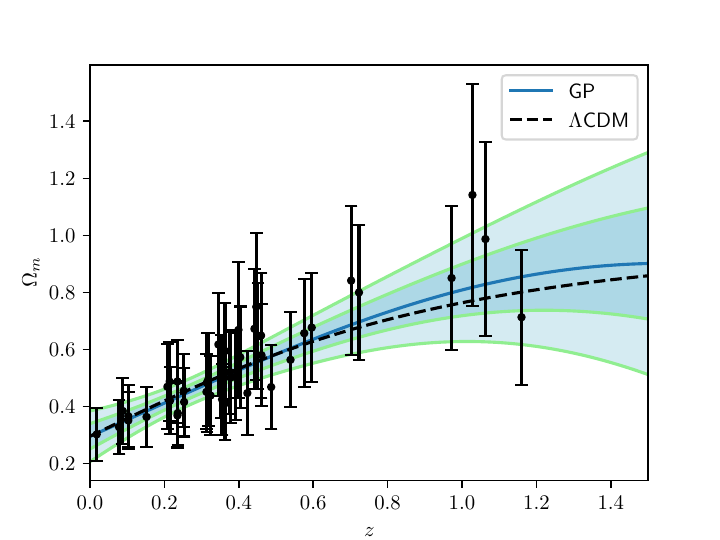}
    \caption{Reconstruction of the matter density parameter $\Omega_m(z)$ from the 44-cluster $f_{\rm gas}$ sample. The shaded regions represent the $1\sigma$ and $2\sigma$ confidence levels, and the dashed line corresponds to the $\Lambda$CDM prediction.}
    \label{fig:44WmGP}
\end{figure}
Rather than performing a fully analytical propagation, we compute the uncertainties numerically using the Python package \texttt{uncertainties}\footnote{\url{https://uncertainties.readthedocs.io/en/latest/}}. {In the Gaussian Process reconstructions of $H(z)$ and $d_L(z)$, we used the full covariance matrices of the observational data. However, in the subsequent error propagation, we did not account for the full covariance matrices of the reconstructed functions. Instead, we propagated the uncertainties independently at each redshift $z$, neglecting correlations between different redshifts. The values of $d_L(z)$ and $H(z)$ at the required redshifts were obtained by linearly interpolating the reconstructions using the \texttt{interp1d} routine from the SciPy package\footnote{\url{https://scipy.org/}}.}

The resulting reconstruction of $\Omega_m(z)$ at $1\sigma$ and $2\sigma$ confidence levels, derived from the 44-cluster sample, is shown in Fig.~\ref{fig:44WmGP}, along with the corresponding inferred values of $\Omega_m(z)$. { The cluster measurements extend up to $z=1.063$ for the 44-cluster sample and $z=1.235$ for the 103-cluster sample. Consequently, the GP curves shown beyond the highest-redshift data point represent extrapolations and are increasingly affected by boundary effects. Our physical interpretation is therefore restricted to the redshift interval directly covered by each cluster sample; the extension to $z=1.5$ is shown only to illustrate the behaviour of the GP reconstruction.}
We find that the theoretical prediction of the $\Lambda$CDM model provides an excellent fit to the entire reconstructed $\Omega_m(z)$ evolution, as highlighted by the dashed line in Fig.~\ref{fig:44WmGP}. This agreement indicates that the data-driven reconstruction is fully consistent with the standard cosmological scenario over the redshift range probed. By extrapolating the reconstruction down to $z = 0$, we obtain an estimate for the present-day matter density parameter, $\Omega_{m0}$. We find
\begin{equation}
\Omega_{m0} = 0.296 \pm 0.044 \quad (1\sigma \ \mathrm{CL}),
\end{equation}
which is in very good agreement with recent cosmological constraints. In particular, our result is consistent with CMB measurements from Planck analyses \cite{Planck:2018vyg}, which typically find $\Omega_{m0} \sim 0.31$, as well as with recent large-scale structure probes such as cosmic shear and galaxy clustering analyses (e.g., DES, KiDS, and HSC), which generally report values in the range $\Omega_{m0} \sim 0.26$--$0.30$, depending on the dataset and analysis methodology \cite{DES:2026mkc,Wright:2025xka,Novaes:2024dyh}. It is important to note that this compatibility is driven by the large uncertainties in our inferences, which stem from the minimal cosmological assumptions underlying our analysis.

Our non-parametric reconstruction based on galaxy cluster data provides a complementary determination of $\Omega_{m0}$, reinforcing the overall concordance of current cosmological observations while remaining fully consistent, within uncertainties, with both early- and late-Universe probes.

\subsection{Sample of 103 clusters}

The second dataset consists of a compilation of 103 gas mass fraction measurements spanning the redshift range $0.047 \le z \le 1.235$, in which $f_{\rm gas}$ is evaluated within the larger radius $r_{500c}$. This sample was assembled by Corasaniti et al.~\cite{corasaniti2021cosmological} by combining several independent datasets, including low-redshift clusters ($z < 0.1$) from the X-COP project \cite{eckert2019non}, intermediate-redshift observations ($0.1 \le z \le 0.3$) \cite{ettori2010mass}, and high-redshift systems ($0.4 \le z \le 1.2$) \cite{ghirardini2017evolution}. 

In contrast to the shell-based approach adopted for the 44-cluster sample, these measurements correspond to integrated quantities computed from the cluster centre out to $r_{500c}$. As a result, they probe a larger fraction of the cluster volume, reducing statistical noise but increasing sensitivity to baryonic physics and mass calibration systematics. The dataset used in this analysis is shown in the right panel of Fig.~\ref{fig:44cldata}.

\begin{figure*}[t]
    \centering
    \includegraphics[width=0.49\textwidth]{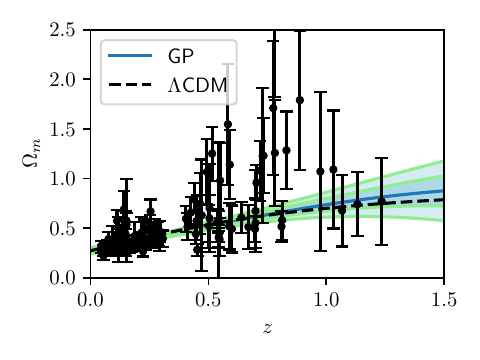}
    \includegraphics[width=0.49\textwidth]{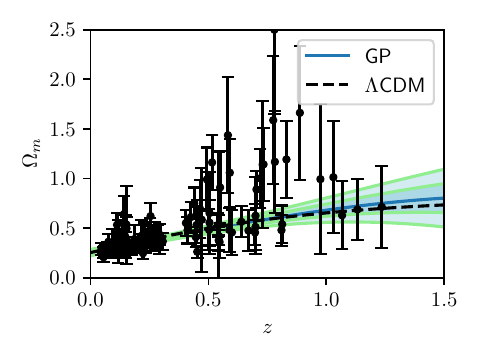}
    \includegraphics[width=0.49\textwidth]{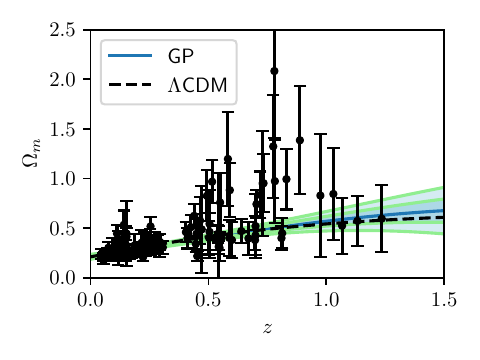}
\caption{Reconstruction of $\Omega_m(z)$ from the 103-cluster $f_{\rm gas}$ sample for different mass calibration schemes. Top left: $K^{\mathrm{CCCP}}$. Top right: $K^{\mathrm{CLASH}}$. Bottom: $K^{\mathrm{CMB}}$.}
\label{fig:103WmGP}
\end{figure*}

\begin{figure}[t]
    \centering
    \includegraphics[width=.49\textwidth]{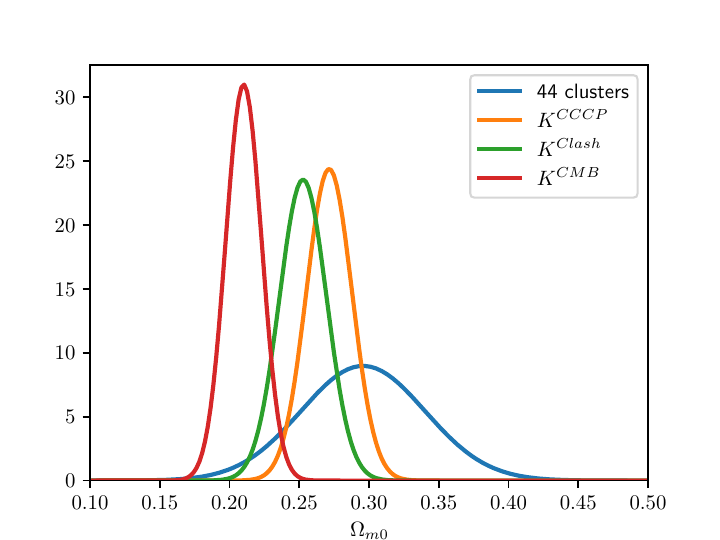}
\caption{{One-dimensional distributions of $\Omega_{m0}$ derived from GP reconstructions of $\Omega_m(z)$, highlighting the impact of different cluster samples and mass calibration schemes.}}
\label{fig:Wm0Likes}
\end{figure}

To model the {baryon depletion factor} for this sample, we adopt a compact polynomial approximation calibrated on the FABLE hydrodynamical simulations \cite{henden2020baryon},
\begin{equation}
\Upsilon(z) = 0.931\,(1 + 0.017z + 0.003 z^2), \qquad \sigma_\Upsilon(z) = 0.04.
\label{eq:gamma_poly}
\end{equation}

{Unlike the shell-based 44-cluster sample, for which the depletion factor refers to the hot gas component, the FABLE-calibrated depletion factor adopted here is defined with respect to the total baryonic content enclosed within $r_{500c}$. Consequently, for this cluster sample, the left-hand side of Eq.~(\ref{eq:f_def}) must be interpreted as the total baryon fraction, $f_b$, rather than the gas mass fraction alone. Therefore, we replace
\begin{equation}
f_{\rm gas}\rightarrow f_b=f_{\rm gas}+f_\star,
\end{equation}
where we adopt a stellar mass fraction $f_\star=0.015\pm0.0015$, following Ref.~\cite{corasaniti2021cosmological}. Since the X-ray observations directly measure $f_{\rm gas}$ rather than $f_b$, we rewrite the above relation as $f_{\rm gas}=f_b-f_\star$. Accordingly, in the numerical implementation, the stellar mass fraction is subtracted from the right-hand side of Eq.~(\ref{eq:f_def}) so that the observed gas mass fraction is used while remaining fully consistent with the definition of $\Upsilon(z)$ adopted in the FABLE simulations. As in the reference analysis, this correction is assumed to be constant over the mass and redshift ranges covered by the sample.}

For the mass calibration parameter $K$, we consider three independent determinations that reflect different sources of systematic uncertainty:

\begin{itemize}
    \item CLASH calibration: accounts for temperature-dependent systematics between \textit{Chandra} and \textit{XMM-Newton} observations, which can bias hydrostatic mass estimates. The CLASH program observed 25 massive clusters ($0.18 < z < 0.89$), enabling precise lensing-based mass reconstructions \cite{Postman2012,Umetsu2016}.
    
    \item CCCP calibration: based on the Canadian Cluster Comparison Project, which combines weak-lensing mass measurements with improved photometric redshift estimates for a sample of $\sim 50$ clusters in the range $0.15 < z < 0.55$ \cite{Hoekstra2012,Mahdavi2013}.
    
    \item CMB-based calibration: obtained from the joint analysis of Sunyaev--Zel'dovich cluster counts and the thermal gas power spectrum measured by \textit{Planck} and the South Pole Telescope (SPT), designed to break degeneracies between cosmology and cluster physics \cite{salvati2018constraints}.
\end{itemize}

These calibrations correspond to
\begin{equation}
\begin{aligned}
K_{\mathrm{CCCP}} &= 0.84 \pm 0.04, \\
K_{\mathrm{CLASH}} &= 0.78 \pm 0.09, \\
K_{\mathrm{CMB}}   &= 0.65 \pm 0.04.
\end{aligned}
\label{Kcons}
\end{equation}

Each adopted value of $K$ therefore defines an independent realization of the cosmological analysis, allowing us to assess the impact of mass calibration systematics on the inferred $\Omega_m(z)$ evolution.

{ To derive $\Omega_m(z)$ from the $f_{\rm gas}(z)$ measurements, we again use Eq.~\eqref{eq:Omega_mz}, adopting $\Upsilon(z)$ and $K$ from Eqs.~\eqref{eq:gamma_poly} and \eqref{Kcons}. As commented earlier, for this sample, the gas mass fraction entering Eq.~\eqref{eq:Omega_mz} is replaced by the total baryon fraction, $f_{\rm b}=f_{\rm gas}+f_{\star}$, consistently with the definition of the depletion factor $\Upsilon(z)$ adopted from the FABLE simulations. The uncertainties associated with $\Upsilon(z)$, the mass calibration parameter $K$, and the stellar mass fraction $f_{\star}$ are propagated through Eq.~\eqref{eq:Omega_mz}, assuming a constant scatter $\sigma_{\Upsilon}=0.04$ for the depletion factor and Gaussian uncertainties for $K$ and $f_{\star}$. The resulting $\Omega_m(z)$ measurements derived from the 103-cluster sample, together with their GP reconstructions, are shown in Fig.~\ref{fig:103WmGP}.}

From the GP reconstruction of $\Omega_m(z)$, we infer the present-day matter density parameter $\Omega_{m0}$, {by evaluating the reconstruction at $z=0$. The corresponding probability density functions} are shown in Fig.~\ref{fig:Wm0Likes}, while the constraints are summarized in Table~\ref{tab:wm0}.

\begin{table}[t]
\centering
\setlength{\tabcolsep}{12pt}
\begin{tabular}{l c}
\hline\hline
Sample & $\Omega_{m0}$ \\
\hline
44 clusters & $0.296 \pm 0.044$ \\
103 clusters ($K^{\mathrm{CCCP}}$) & $0.271 \pm 0.016$ \\
103 clusters ($K^{\mathrm{CLASH}}$) & $0.253 \pm 0.017$ \\
103 clusters ($K^{\mathrm{CMB}}$) & $0.210 \pm 0.013$ \\
\hline
\end{tabular}
\caption{Constraints on the present-day matter density parameter $\Omega_{m0}$ derived from GP reconstructions of $\Omega_m(z)$. The table reports results for the 44-cluster sample and for the 103-cluster compilation, highlighting the impact of different mass calibration schemes ($K^{\mathrm{CCCP}}$, $K^{\mathrm{CLASH}}$, and $K^{\mathrm{CMB}}$) on the inferred values.}
\label{tab:wm0}
\end{table}
{ The spread among the values reported in Table~I is significantly larger than the quoted statistical uncertainties associated with each individual reconstruction. This indicates that, within the present analysis, the dominant source of uncertainty in the determination of $\Omega_{m0}$ is the adopted cluster mass calibration rather than the statistical precision of the GP reconstruction. Therefore, the values listed in Table~I should not be interpreted as independent high-precision measurements of $\Omega_{m0}$, but rather as the cosmological inferences obtained under different, physically motivated cluster mass calibration schemes.}

Moreover, an important aspect of the results presented in Table~\ref{tab:wm0} is that the inferred values of $\Omega_{m0}$ can be directly connected to the well-known degeneracies involving $H_0$ and the clustering amplitude parameter $S_8$. In particular, variations in $\Omega_{m0}$ are expected to correlate with shifts in both the expansion rate and the growth of cosmic structure.

From a large-scale structure perspective, lower values of $\Omega_{m0}$ generally lead to a suppression of structure growth, resulting in smaller values of $S_8$. In this context, the results obtained using the CCCP and CLASH calibrations, $\Omega_{m0} = 0.271 \pm 0.016$ and $0.253 \pm 0.017$, respectively, naturally map onto the region of parameter space favoured by cosmic shear surveys (e.g., DES, KiDS, HSC) \cite{DES:2026mkc,Wright:2025xka,Novaes:2024dyh}, which consistently prefer lower values of $S_8$ compared to Planck CMB results \cite{Planck:2018vyg}. {  Therefore, these calibrations are qualitatively consistent with scenarios associated with a lower $S_8$.}

{ On the other hand, the CMB-based calibration yields a lower value, $\Omega_{m0} = 0.210 \pm 0.013$, which would be associated with an even stronger suppression of structure formation.} If combined with standard assumptions for $\sigma_8$, this would correspond to values of $S_8$ well below those inferred from current weak lensing surveys, potentially overcorrecting the existing tension. This suggests that such a low $\Omega_{m0}$ may be difficult to reconcile simultaneously with both CMB and large-scale structure observations, unless accompanied by additional modifications to the growth history. In \cite{Ruiz-Zapatero:2022zpx}, using a combination of geometrical and growth probes, the authors find $\Omega_{m0} = 0.224 \pm 0.066$, which is lower but still statistically compatible with the Planck 2018 cosmology. Notably, this result is remarkably consistent with our estimate $\Omega_{m0} = 0.210 \pm 0.013$ obtained under the CMB-based calibration, reinforcing the interpretation that such low values of $\Omega_{m0}$ are closely linked to calibration choices and residual systematics in cluster-based analyses.
{ From the perspective of the $H_0$ tension, the situation is more subtle. In the context of $\Lambda$CDM and several late-time models, a lower $\Omega_{m0}$ is typically associated with a higher inferred value of $H_0$ when fitting background observables, due to parameter degeneracies in distance measurements. Therefore, the trend observed across the different calibrations---particularly the low value obtained with the CMB-based calibration---may qualitatively point in the direction preferred by local distance ladder measurements of $H_0$.

However, it is important to emphasize that the present analysis does not directly constrain either $H_0$ or $S_8$, and therefore the above discussion should be regarded as a qualitative interpretation of the inferred $\Omega_{m0}$ values. Moreover, simultaneously resolving both the $H_0$ and $S_8$ tensions within $\Lambda$CDM is known to be challenging, as changes that alleviate one tension often exacerbate the other (see \cite{CosmoVerseNetwork:2025alb} for a review).}

Taken together, these results highlight that cluster-based determinations of $\Omega_{m0}$ occupy a critical position in the current landscape of cosmological tensions. { The CCCP and CLASH calibrations yield values that are qualitatively consistent with those preferred by current large-scale structure observations, often associated with a ``low-$S_8$'' Universe, while remaining compatible with standard CMB constraints within uncertainties. In contrast, the CMB-calibrated case may point toward a more extreme scenario, potentially indicating either residual systematics in the mass calibration or the need for extensions beyond the standard cosmological model.} In this context, as emphasized by \cite{Pedrotti:2024kpn}, the Hubble tension is intrinsically multidimensional: once $\Omega_{m0}$ and $\omega_{b0}$ are calibrated, increasing $H_0$ requires a higher $\omega_{c0}$ and typically a larger $S_8$. This further underscores the importance of robust and independent constraints on $\Omega_{m0}$, as provided by cluster-based analyses.

Overall, the dependence of $\Omega_{m0}$ on the calibration parameter $K$ reinforces the conclusion that astrophysical systematics remain the dominant limitation in cluster cosmology. At the same time, it underscores the potential of $f_{\rm gas}$ measurements, when properly calibrated, to provide valuable insight into the interplay between background expansion and structure growth, and to contribute to the ongoing discussion of cosmological tensions.

In general, the reconstructed $\Omega_m(z)$ evolution obtained from both cluster samples shows no statistically significant deviation from the behaviour expected in the standard $\Lambda$CDM framework over the redshift range probed by the data, corresponding to approximately 10 billion years of cosmic evolution. Although individual measurements exhibit substantial scatter—primarily driven by uncertainties in mass calibration and in the modelling of the gas depletion factor—the Gaussian Process reconstructions remain consistent with the standard scaling $\rho_m \propto (1+z)^3$ within current uncertainties.

The larger 103-cluster compilation provides significantly tighter constraints compared to the 44-cluster sample, reflecting its improved statistical power. Nevertheless, both datasets are mutually consistent within their respective uncertainties. We emphasize, however, that the inferred value of $\Omega_{m0}$ depends sensitively on the adopted calibration of the mass bias parameter $K$, highlighting the dominant role of cluster-related systematics in cosmological analyses based on $f_{\rm gas}$ measurements.

Despite these limitations, a key strength of the present approach is that it enables the reconstruction of $\Omega_m(z)$ and the estimation of $\Omega_{m0}$ {in a minimally model dependent approach}, without assuming a specific cosmological background or imposing a parametric form for the evolution of the matter density.  {We verified that varying the adopted baryon-density prior within its $1\sigma$ uncertainty produces negligible changes in the reconstructed $\Omega_m(z)$ compared with the statistical uncertainties and the mass-calibration systematic.}

\section{\label{sec:conclusion}Conclusion}

In this work, we have investigated the redshift evolution of the matter density parameter, $\Omega_m(z)$, by combining galaxy cluster gas mass fraction measurements with cosmic chronometer $H(z)$ data and type Ia supernova luminosity distances. Our analysis provides a complementary probe of the normalization of the cosmological matter sector. 
The analysis was performed using two independent cluster samples comprising 44 and 103 systems. Our methodology relies on Gaussian Process regression, enabling a non-parametric reconstruction of $\Omega_m(z)$.
 
This approach enables a comparison between the reconstructed $\Omega_m(z)$ and the evolution implied by the standard matter-density scaling $\rho_m \propto (1+z)^3$ predicted by the $\Lambda$CDM framework, while relying only on a minimal set of well-motivated cosmological assumptions.

From the reconstructed evolution, we obtained constraints on the present-day matter density parameter. For the 44-cluster sample, we find $\Omega_{m0}=0.296 \pm 0.044$. For the 103-cluster compilation, we obtain $\Omega_{m0}=0.271 \pm 0.016$, $\Omega_{m0}=0.253 \pm 0.017$, and $\Omega_{m0}=0.210 \pm 0.013$, depending on the adopted calibration of the mass bias parameter $K$. While the normalization of $\Omega_{m0}$ exhibits a clear dependence on the adopted calibration scheme, the reconstructed $\Omega_m(z)$ evolution remains consistent, within current uncertainties, with the expectations of the $\Lambda$CDM model over the redshift range probed by the data, corresponding to approximately 10 Gyr of cosmic history.

From a broader cosmological perspective, the inferred values of $\Omega_{m0}$ provide an interesting comparison with the current landscape of cosmological constraints. The CCCP and CLASH calibrations yield values that are qualitatively consistent with those preferred by some large-scale structure probes, often associated with a ``low-$S_8$'' Universe, while remaining compatible with CMB constraints within the current uncertainties. In contrast, the CMB-based calibration favours a lower value of $\Omega_{m0}$. {However, these differences should be interpreted with caution, as the present error budget is dominated by systematic uncertainties associated with the mass calibration. Therefore, the results do not by themselves provide robust evidence for a preference between these calibrations or for departures from the standard cosmological framework. Instead, they highlight the importance of improved and independent mass calibrations for obtaining more precise constraints on $\Omega_{m0}$ and, consequently, on the interpretation of cosmological tensions.}

{In this context, and as discussed in the literature, the Hubble tension is intrinsically multidimensional: once $\Omega_{m0}$ and $\omega_{b0}$ are calibrated, increasing $H_0$ generally requires a higher $\omega_{c0}$ and a larger clustering amplitude $S_8$. Our results therefore emphasize that independent constraints on $\Omega_{m0}$, together with improved control of mass-calibration systematics, are essential for assessing the extent to which cluster-based measurements can contribute to the interpretation of current cosmological tensions.}

Overall, our results demonstrate that cluster-based measurements of $f_{\rm gas}$, when combined with non-parametric reconstruction techniques, constitute a complementary probe of the evolution of the matter density. While the normalization of $\Omega_{m0}$ remains primarily limited by mass calibration systematics, the reconstructed $\Omega_m(z)$ evolution is consistent with the standard cosmological framework within the current observational uncertainties. Looking ahead, upcoming X-ray and Sunyaev--Zel'dovich surveys, together with improved weak-lensing mass calibrations (e.g., \cite{CMB-S4:2024zgz,HSC:2025lum,Shin:2025jxz,Bocquet:2026lxi,Cerardi:2025dye}), are expected to substantially mitigate these systematics, enabling cluster cosmology to play an increasingly significant role in precision tests of $\Lambda$CDM and in probing potential extensions beyond the standard model.
\\

\begin{acknowledgments}
{We thank the referees for the careful reading of the manuscript and for several constructive comments and suggestions.}
This study was financed in part by the Coordena\c{c}\~ao de Aperfei\c{c}oamento de Pessoal de N\'ivel Superior - Brasil (CAPES) - Finance Code 001. RFLH acknowledges financial support from Conselho Nacional de Desenvolvimento Cient\'ifico e Tecnol\'ogico (CNPq). JFJ acknowledges financial support from Conselho Nacional de Desenvolvimento Cient\'ifico e Tecnol\'ogico (CNPq) (No. 314028/2023-4). R.C.N. acknowledges financial support from the CNPq through Grant No. 304306/2022-3, and partial financial support from the Fundação de Amparo à Pesquisa do Estado do Rio Grande do Sul (FAPERGS, Research Support Foundation of the State of Rio Grande do Sul) through Grant No. 23/2551-0000848-3 and Grant No. 25/2551-0002612-1. 
\end{acknowledgments}

\bibliography{references}

\end{document}